\documentclass[a4paper,fleqn,usenatbib]{mnras}

\usepackage{newtxtext,newtxmath}

\usepackage[T1]{fontenc}
\usepackage{ae,aecompl}

\usepackage{graphicx}	
\usepackage{amsmath}	

\usepackage{amssymb}	

\title[Pseudobulges in S0 galaxies]{Age bimodality in the central region of pseudobulges in S0 galaxies}

\author[P. K. Mishra et al.]{
Preetish K. Mishra,$^{1}$\thanks{E-mail: preetish@ncra.tifr.res.in}
Sudhanshu Barway,$^{2}$\thanks{E-mail: barway@saao.ac.za }
Yogesh Wadadekar$^{1}$\thanks{E-mail: yogesh@ncra.tifr.res.in}
\\
$^{1}$National Centre for Radio Astrophysics, TIFR, Post Bag 3, Ganeshkhind, Pune 411007, India\\
$^{2}$South African Astronomical Observatory, P.O. Box 9, 7935,
Observatory, Cape Town, South Africa
}
\date{Accepted for publication in MNRAS Letters}

\pubyear{2017}

\begin{document}
\label{firstpage}
\pagerange{\pageref{firstpage}--\pageref{lastpage}}
\maketitle

\begin{abstract}

We present evidence for bimodal stellar age distribution of
pseudobulges of S0 galaxies as probed by the $D_n(4000)$ index.  We do
not observe any bimodality in age distribution for pseudobulges in
spiral galaxies. Our sample is flux limited and contains 2067 S0 and
2630 spiral galaxies drawn from the Sloan Digital Sky Survey. We
identify pseudobulges in S0 and spiral galaxies, based on the position
of the bulge on the Kormendy diagram and their central velocity
dispersion. Dividing the pseudobulges of S0 galaxies into those
containing old and young stellar populations, we study the connection
between global star formation and pseudobulge age on the $u-r$
color-mass diagram. We find that most old pseudobulges are hosted by
passive galaxies while majority of young bulges are hosted by galaxies
which are star forming. Dividing our sample of S0 galaxies into
early-type S0s and S0/a galaxies, we find that old pseudobulges are
mainly hosted by early-type S0 galaxies while most of the pseudobulges
in S0/a galaxies are young. We speculate that morphology plays a
strong role in quenching of star formation in the disc of these S0
galaxies, which stops the growth of pseudobulges, giving rise to old
pseudobulges and the observed age bimodality.

\end{abstract}




\section{Introduction}
\label{sec:intro}
Recent advances in our knowledge of galaxy structure have shown that
the central spheroidal component of galaxies, i.e. the galaxy bulge,
comes in two flavours : classical and pseudo.  Classical bulges are
formed by fast and violent processes such as major mergers or by
sinking and coalescence of giant gas clumps found in high redshift
discs \citep{Elmegreen2008,Kormendy2016}. Pseudobulges, on the other
hand, are thought to be formed by the slow rearrangement of gaseous
material from the disc to the central region of galaxies. The process
forming the pseudobulge can either be totally internal in nature,
driven by non-axisymmetric structures such as bars, ovals
etc. \citep{Kormendy2004}, or can involve external processes such as
minor mergers \citep{Eliche-Moral2011}. The different formation
mechanisms, for the two bulge types, leave their imprint on the
stellar population of these bulges. Previous studies have associated
formation of classical bulge with rapid and efficient star formation,
while the pseudobulges are formed slowly at lower redshift
\citep{Sanchez2016}. As as result, the stellar populations of
pseudobulges are found to be younger, on average, as compared to those
of classical bulges \citep{Gadotti2009}.  \par

 S0 galaxies are an intermediate transition class between elliptical
 and spiral galaxies on the Hubble tuning fork diagram
 \citep{Hubble1936}. They are thought to have formed from  spiral
 galaxies via different physical processeses such as major/minor
 mergers \citep{Quereteja2015} or due to fading of spiral arms and the
 quenching of star formation in the disc through environmental processes such as tidal
stripping and galaxy harassment \citep{Moore1996}, starvation \citep{Bekki2002} etc. By studying the stellar
population of these galaxy bulges, we can try to understand how
morphological transformation of a spiral into an S0 galaxy, affects
the properties of the bulge. One can put constraints on
 the formation channel of S0 galaxies, by knowing the type of bulge
 that they host.  For example, the presence of a pseudobulge in a S0
 galaxy helps us to discard the major merger driven formation channel
 for that particular galaxy. One  expects that the formation of an
 S0 galaxy due to the quenching of star formation in the disc of their
 progenitor spirals must leave an imprint on the star formation
 history of pseudobulges which are formed from these discs. By comparing
 the properties of the stellar population of pseudobulges hosted by S0
 galaxies and the ones hosted by spirals, one can try to understand
 the impact of morphological transformation on the properties of
 pseudobulges.  \par
 
 In this letter, we study the stellar population of
 pseudobulges hosted in S0 and spiral galaxies. The organization of
 this letter is as follows, Section \ref{sec:data} describes our data
 and sample selection. Section \ref{sec:result} describes our results
 and discusses the findings before summarising the results in Section
 \ref{sec:sum}. Throughout this work, we have used the WMAP9
 cosmological parameters: $H_0$ = 69.3 km s$^{-1}$Mpc$^{-1}$,
 $\Omega_m$= 0.287 and $\Omega_{\lambda}$= 0.713.

\section{Data and sample selection}
\label{sec:data}

In order to construct a statistically significant sample of S0
galaxies for our study, we started with data provided in
\cite{Nair2010}, which is a catalogue of detailed visual
classification for nearly 14,000 spectroscopically targeted galaxies
in the SDSS DR4. The \cite{Nair2010} catalogue is a flux limited
sample with an extinction corrected limit of $g < 16$ mag in the SDSS
$g$ band, spanning the redshift range $0.01 < z < 0.1$. This catalogue
provides information on the morphological T type and other
morphological features such as a bar, ring etc. In addition to
information on morphology, it also lists the stellar mass of each
galaxy as estimated by \cite{Kauffmann2003}, and group membership
information from the \cite{Yang2007} catalogue. In order to obtain
information on the structural components of these galaxies, we cross
matched \cite{Nair2010} catalogue with the data provided in
\cite{Simard2011} catalogue. \cite{Simard2011} provides us with
two-dimensional, point-spread-function-convolved, bulge+disc
decompositions in the $g$ and $r$ bands for a sample of 1,123,718
galaxies from the SDSS Data Release 7 \citep{Abazajian2009}. The cross
match resulted in 12,063 galaxies, which we refer to as the parent
sample hereafter.

 \par \cite{Simard2011} have fitted each galaxy in their sample with
 three different light profile models: a pure S\'ersic model, an
 $n_b$ = 4 bulge + disc model, and a S\'ersic (free $n_b$) bulge +
 disc model. One can choose the most appropriate model for a given
 galaxy using the F-test probability criteria. For our study, we have
 chosen only those galaxies from our parent sample where a bulge + disc
 model is preferred over a single S\'ersic model. We chose the free
 $n_b$ bulge + disc model for the disc galaxies in our sample as previous studies have shown that the bulges of S0s and spirals span a wide range of values of S\'ersic index \citep{Balcells2007, Laurikainen2010}. To find the appropriate model for the ellipticals in our sample we have carried out a comparison between the two available bulge + disc models and have found that majority of ellipticals are better fitted with $n_b$ = 4 bulge + disc model as compared to a free $n_b$ + disc model. Therefore, we use $n_b$ = 4 bulge + disc model to obtain relevant structural parameters of elliptical galaxies in our parent sample.\par 
 
The allowed range of bulge S\'ersic index ($n_{b}$) in the
 \cite{Simard2011} sample is $0.5<n_{b}<8$. From the literature it has been
 known that the high values of the  S\'ersic  index are often associated
 with fitting problems \citep{Meert2015}. In our sample, we find that
 the mean error in bulge S\'ersic index for the galaxies having
 $n_{b}>7.95$ is around twice the error in S\'ersic index estimate
 below $n_{b} = 7.95$. \cite{Simard2011} also report that a
 significant number galaxies with these high value of $n_{b}$, contain
 a nuclear source or a bar + point like source. Presence of such sources
 might affect the reliability of bulge parameter estimation at these
 high values of $n_{b}$, hence we have excluded galaxies having
 $n_{b}\geqslant 7.95$ from our parent sample. To further enhance the
 quality of our sample, we impose a further selection cut based on the
 error in estimation of $n_{b}$, in which we demand that no galaxy in
 our parent sample should have error in $n_{b}$ which is greater than
 the mean + one sigma of error distribution. We apply a final selection cut in our sample, in which we remove all the galaxies which host a bar. Since, \cite{Simard2011} does not fit for the bar profile in their decomposition, there is a chance that estimated bulge light profile might be contaminated by the bar light profile and one might be over estimating the bulge sizes etc. We have used flags provided in \cite{Nair2010} to identify and remove all galaxies which contain a bar from our sample. 
 
 \par Application of these cuts on the parent sample resulted in a
 final sample of 1742 elliptical and 4697 disc galaxies which are
 modelled by a $n_b = 4$ bulge + disc and free $n_b$ bulge + disc
 galaxy models respectively. The mean error on $n_b$ in our final
 sample of disc galaxies is 0.17. Out of these 4697 disc galaxies,
 2067 of them are S0s and 2630 are spiral galaxies. To study the
 stellar population of these disc galaxies in our final sample, we
 have obtained relevant measurement of the $D_n$(4000) index ($d4000_n$; as
 defined in \citealt{Balogh1999}) from the table {\it galSpecIndx} using
 the SDSS DR13 \citep{SDSS2016} CASJobs. The median error on the measured value of $D_n$(4000)is 0.013 for our sample.

\begin{figure}
	\includegraphics[width=0.5\textwidth]{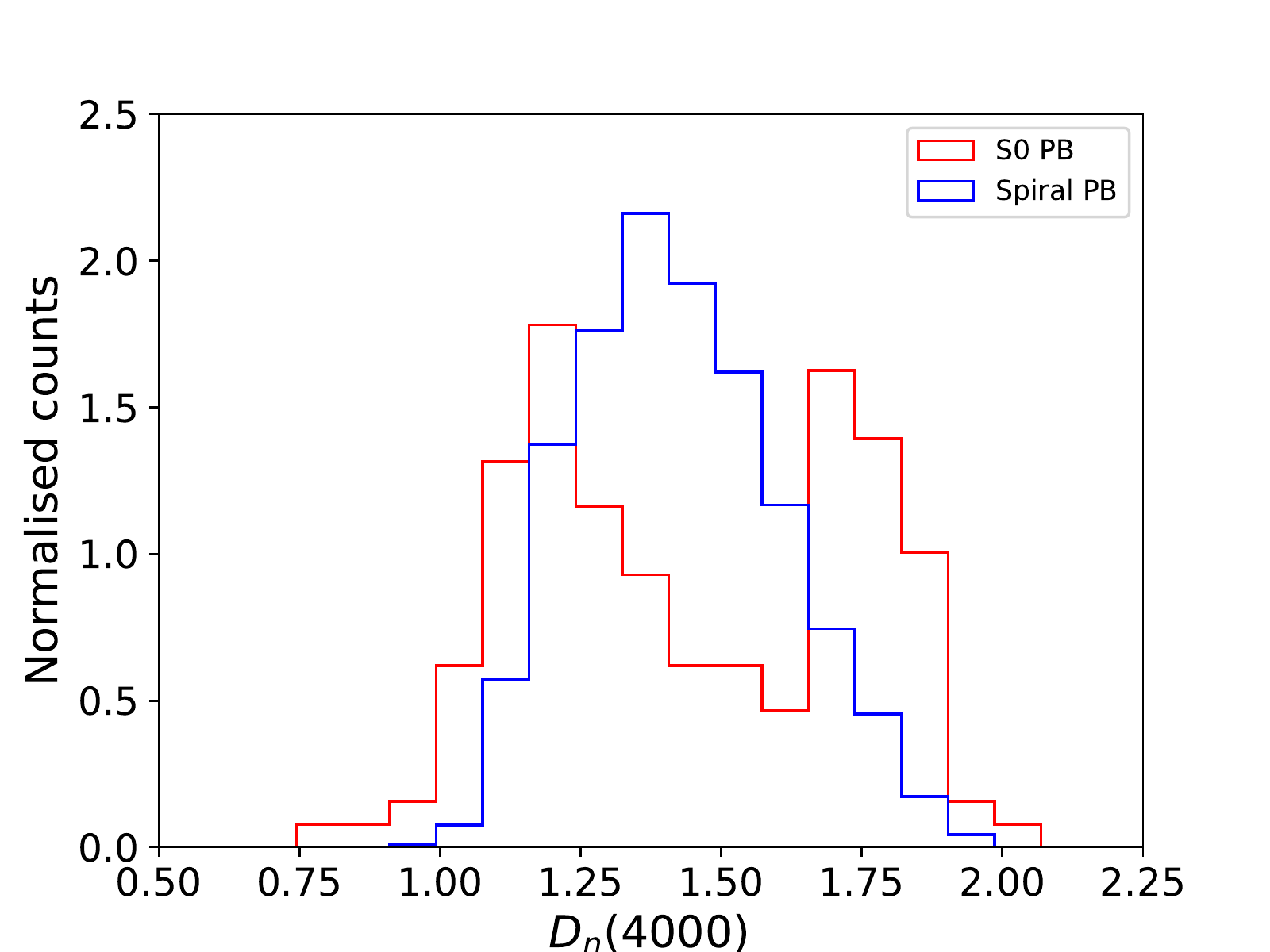}
   \caption{Distribution of D$_n$(4000) index in pseudobulge-hosting S0 (red solid line) and spiral (blue solid) galaxies. The histograms are normalised by the area under the curve.}
   \label{fig:s0sp}
\end{figure} 
 
\begin{figure*}

\includegraphics[width=.33\textwidth]{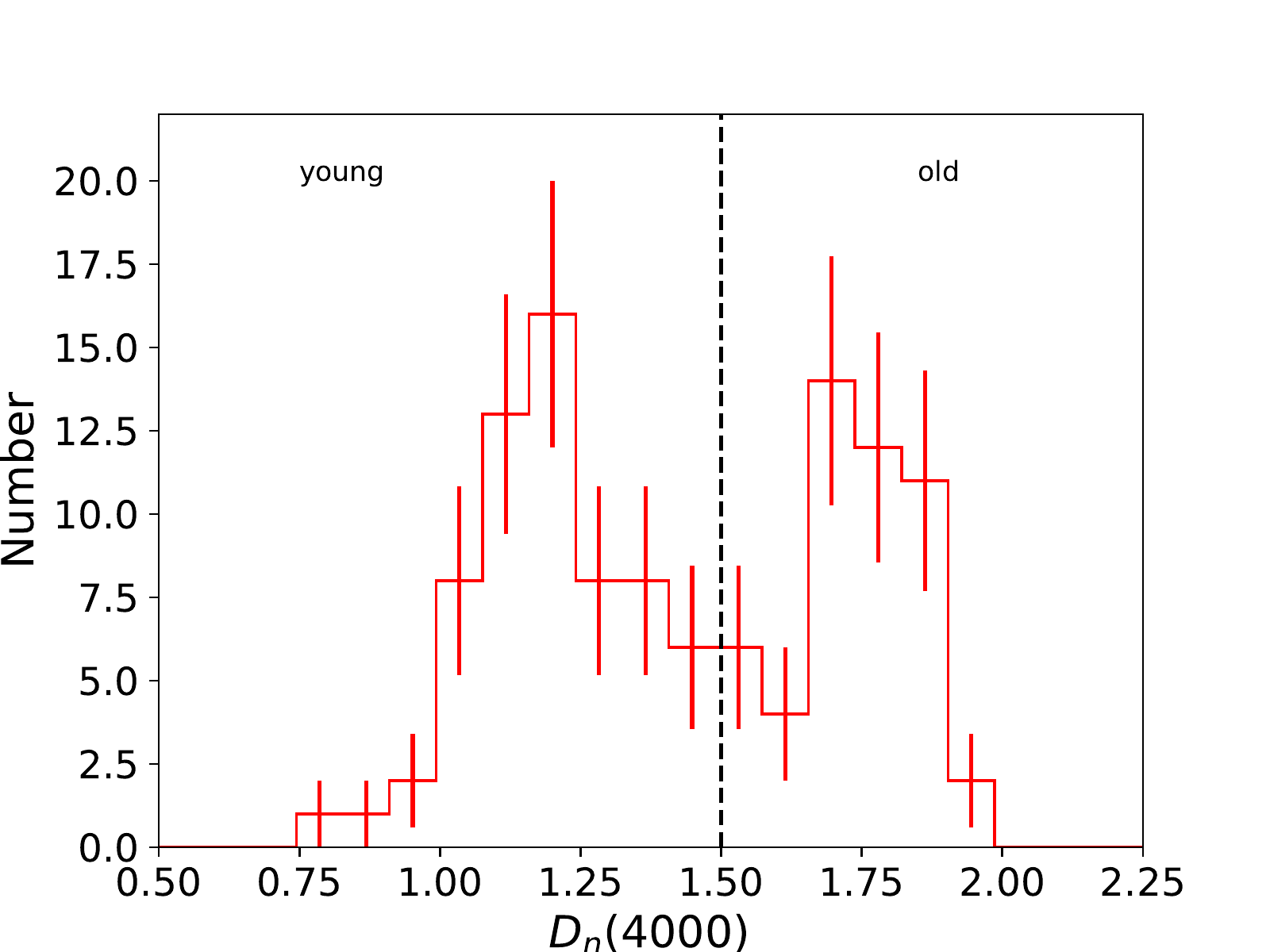}\hspace{-1.5em}
\includegraphics[width=.33\textwidth]{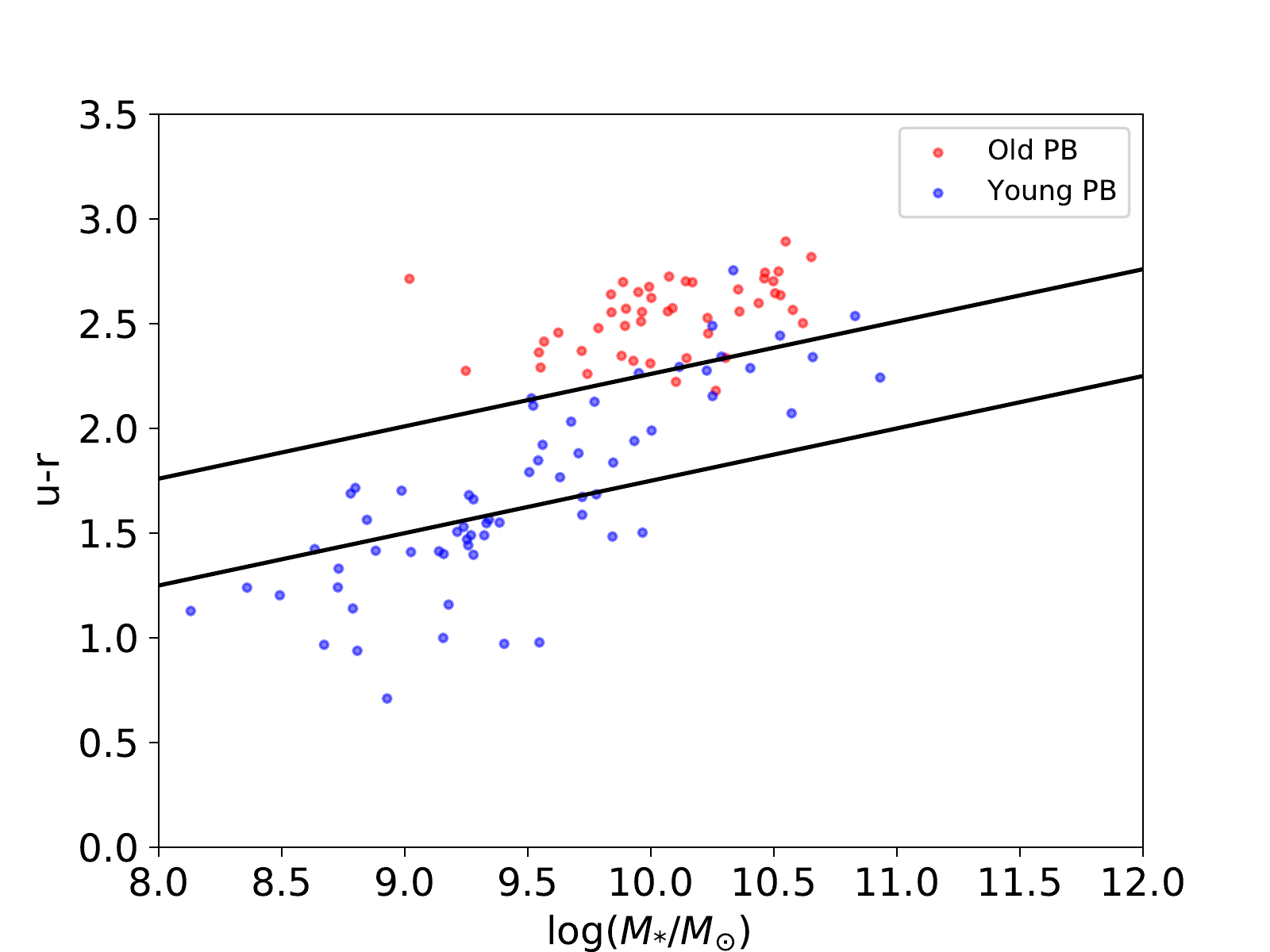}\hspace{-1.5em}
\includegraphics[width=.33\textwidth]{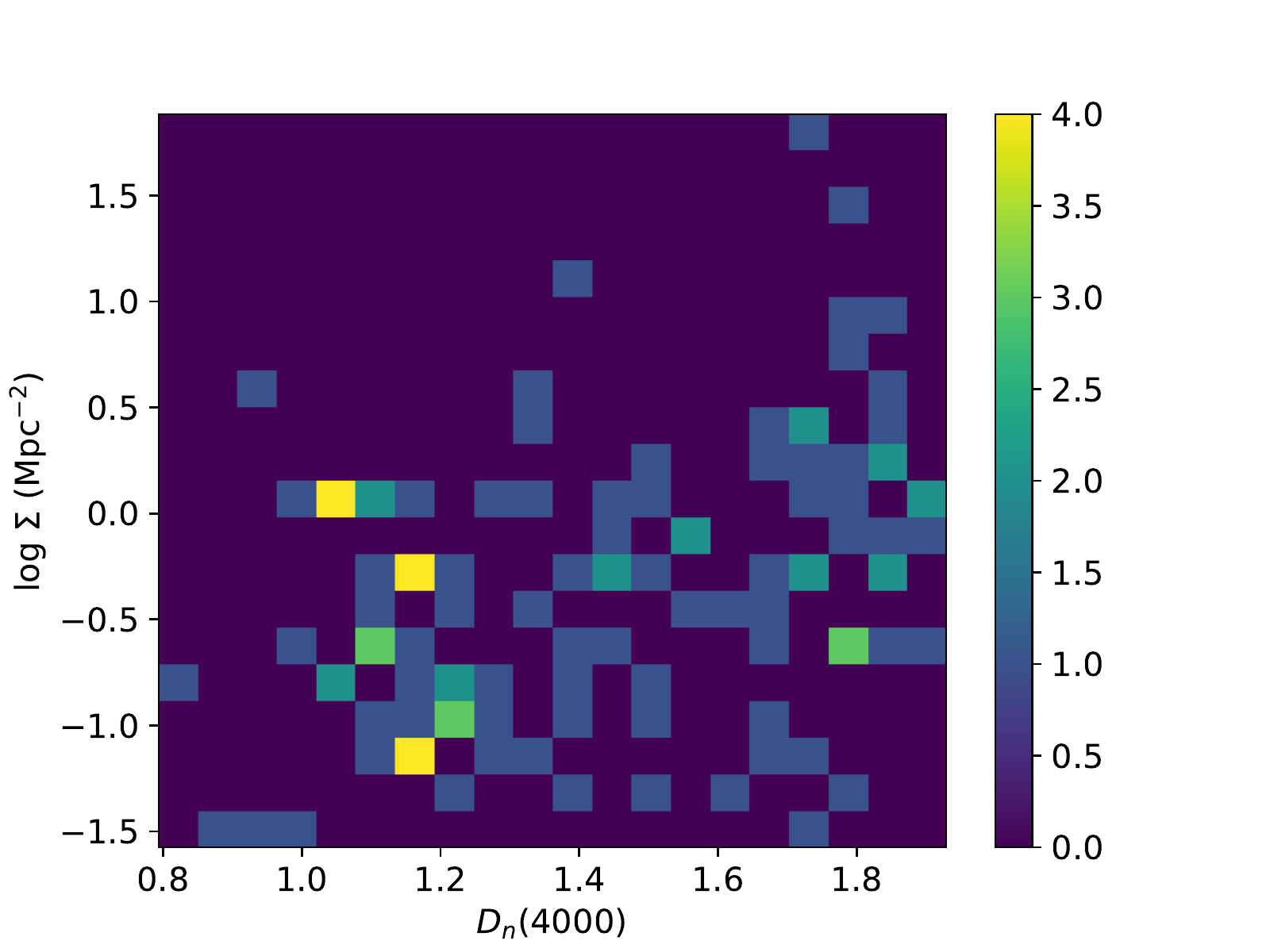}

\caption{\textbf{Left} : $D_n(4000)$ distribution of pseudobulge of S0 galaxies. The error bars on the histogram are Poisson errors and the median error in $D_n(4000)$ measurement is 0.013. The dividing line $D_n(4000)$ = 1.5 separates the pseudobulges into young and old population. \textbf{Middle}: position of pseudobulge-hosting S0 galaxies on u-r colour-mass diagram. Red and blue colours denote old and young pseudobulges respectively. The median errors in mass and colour measurement are 0.15 dex and 0.015 respectively. \textbf{Right}: Two dimensional $D_n(4000)$-environmental density histogram for pseudobulge-hosting S0 galaxies.}
\label{fig:fig1}

\end{figure*}

\section{Results and discussion}

\label{sec:result}
\subsection{Identifying pseudobulges}

We have classified bulges in our sample by combining two independent
criteria for bulge type identification, one of them coming from
photometry and the other coming from
spectroscopy. \cite{Kormendy2016} has shown that the failure
probability of a single criteria can range from 10-20 \%, and the
failure probability goes down significantly if one uses two or more
independent criteria to identify bulges. \par

The photometric criterion for the identification of pseudobulges
follows \cite{Gadotti2009}, which involves classification of bulge
types based on their position on the Kormendy diagram
\citep{Kormendy1977}. This diagram is a plot of the average surface
brightness of the bulge within its effective radius $\langle\mu_b (<
r_e)\rangle$ against the logarithm of the bulge effective radius
$r_{e}$. Elliptical galaxies are known to obey a tight linear relation
on this plot. Classical bulges being structurally similar to
ellipticals obey a similar relation while pseudobulges being
structurally different, lie away from it. Any bulge that deviates more
that three times the r.m.s. scatter from the best fit relation for
ellipticals is classified as pseudobulge by this criterion
\citep{Gadotti2009}. This physically motivated classification scheme
has been used in recent works \citep{Vaghmare2013,
  Mishra2017,Neumann2017}

\par The Kormendy equation was obtained by fitting ellipticals in our final sample using $r$ band data. The equation for the best fit line is \\

$\langle\mu_b (< r_e)\rangle$ = $(2.330 \pm 0.047)$ log($r_e$) + $(18.160 \pm 0.024)$
\\

The rms scatter in $\langle\mu_b (< r_e)\rangle$ around the best fit line is 0.429. All galaxies which lie away more than 3 sigma scatter from this relation are classified as pseudobulge hosts.

\par \cite{Fisher2016} have suggested that if a bulge is found to have
a central velocity dispersion ($\sigma_{0}$) greater than 130 km
s$^{-1}$ , then it is most likely to be a classical bulge. We also
impose this criterion coming from spectroscopic measurements on our
sample, in which we demand that in order to be classified as a
pseudobulge, the central velocity dispersion ($\sigma_{0}$) of the
bulge should be less than 130 km s$^{-1}$.  After the simultaneous
application of these two criteria, we find that 156 (7.5\%) out of
2067 S0 galaxies a host pseudobulge while 1118 (42.5\%) out of 2630
spirals are pseudobulge hosts. All the subsequent analysis presented
in this work has been carried on these pseudobulge-hosting spiral and
S0 galaxies. Previous works have quoted pseudobulge fraction in
spirals and S0s to be 32\% \citep{Gadotti2009} and 14\%
\citep{Vaghmare2013} respectively. Pseudobulges are more
  commonly seen in low mass galaxies \citep{Fisher2016,
    Mishra2017}. In his work \cite{Gadotti2009} has selected spiral
  galaxies having stellar mass greater than $10^{10} M_{\odot}$ which
  is much greater than lower limit of stellar mass ($10^{8}
  M_{\odot}$) in our sample. The sample of \cite{Vaghmare2013} is, on
  average, fainter than ours. The different pseudobulge fraction that
  we obtain as compared to previous works is most likely due to the
  different mass range of our sample.


\begin{figure*}
  \centering
  \begin{minipage}[b]{0.4\textwidth}
    \includegraphics[width=\textwidth]{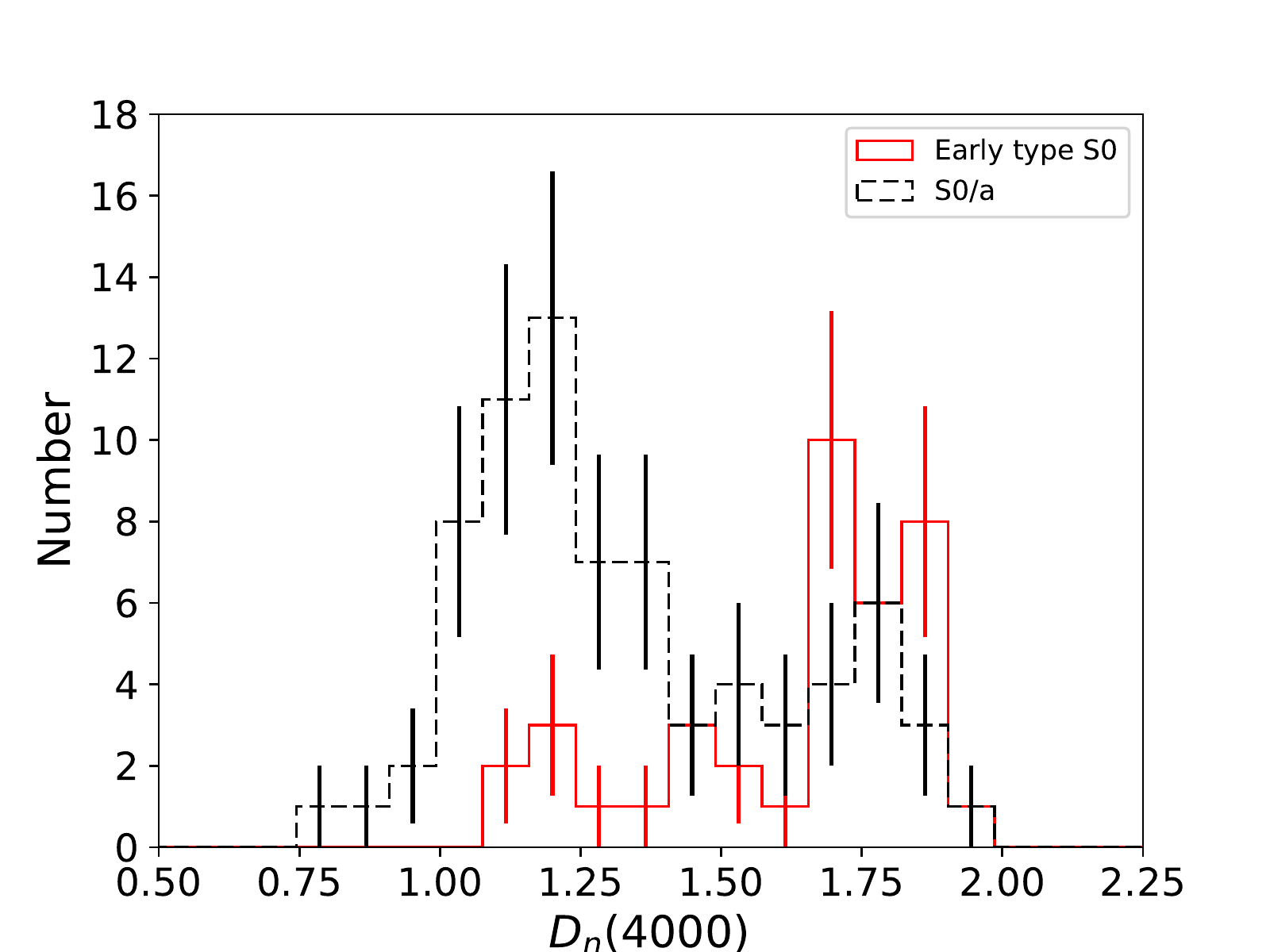}
    
  \end{minipage}
  \hspace{-1.98em}%
  \begin{minipage}[b]{0.4\textwidth}
    \includegraphics[width=\textwidth]{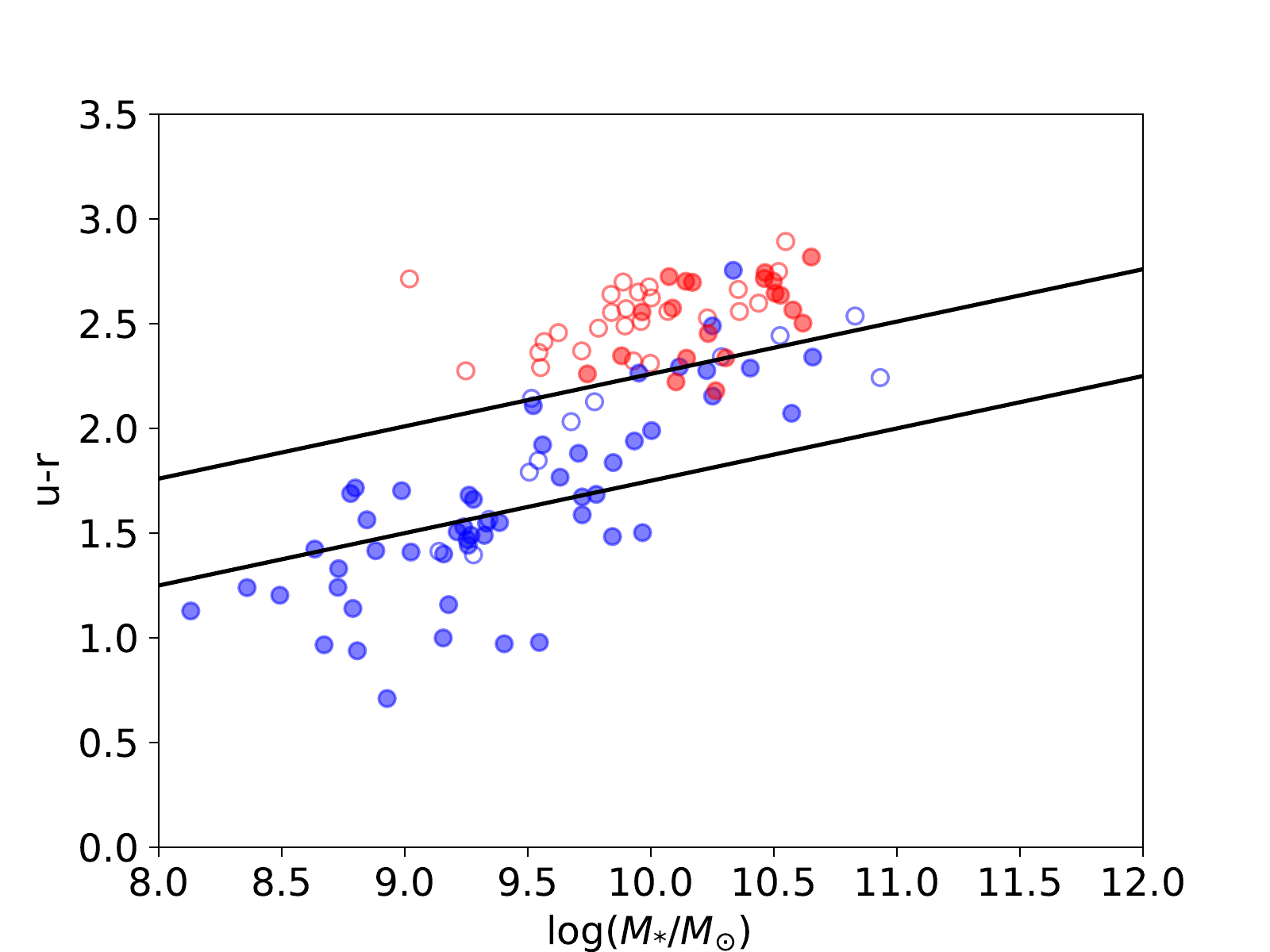}
    
  \end{minipage}
  \caption{\textbf{Left} : $D_n(4000)$ distribution of pseudobulges of early-type S0 and S0/a galaxies. The error bars on both histograms are the Poisson errors. \textbf{Right}: position of pseudobulge-hosting early-type S0 (open circles) and S0/a (filled circle) galaxies on u-r colour mass diagram. Red and blue colours denote old and young pseudobulge hosts respectively. The median errors on $D_n(4000)$, stellar mass and u-r colour are same as in Figure \ref{fig:fig1}}
\label{fig:both}
\end{figure*} 

\subsection{Age bimodality in pseudobulges of S0 galaxies}

In order to study the stellar population of pseudobulges in our
sample, we have made use of available measurement of the 4000 {\AA}
spectral break index ($D_n(4000)$). The strength of the 4000 {\AA}
spectral break arises due to accumulation of absorption lines of
(mainly) metals in the atmosphere of old, low mass stars and by a lack
of hot, blue stars in galaxies. The strength of this break is
quantified by the $D_n(4000)$ index. The $D_n(4000)$ index is a reliable indicator of mean age of galaxy stellar population. In literature, galaxies with break strength of $D_n(4000)$ $\sim 1.3$ and $D_n(4000)$ $\sim 1.8$ have been quoted to have light weighted mean stellar ages of $\sim$ 1-2 Gyr and $\sim 10$ Gyr respectively \citep{Kauffmann2003}. In Figure \ref{fig:s0sp}, we have plotted the
distribution of $D_n(4000)$ index for pseudobulge-hosting S0 and
spiral galaxies. One can clearly see the distribution of $D_n(4000)$
index is bimodal for the pseudobulge-hosting S0 galaxies while spirals
do not show any bimodality in their $D_n(4000)$ distribution. 

The measurement of $D_n(4000)$ index comes from central 3
  arcsec as probed by SDSS fiber aperture. Since the galaxies in our
  sample come in different physical sizes and are distributed over a
  redshift range of $0.01 < z < 0.1$, there exists a possibility that
  the SDSS fiber aperture is not predominantly tracing the bulge
  region and is contaminated by the light from the inner disk. This
  can cause a bias in $D_n(4000)$ measurements towards younger
  ages. In order to correct for this effect, we have chosen to retain
  only those pseudobulge-hosting S0 galaxies which have their light
  profiles such that the bulge is brighter than the disc everywhere
  within the region traced by the SDSS fiber aperture. For each galaxy
  the bulge and disc light profile is obtained from the decompositions
  of \cite{Simard2011}. Out of the original 156 S0 pseudobulges, 112
  satisfied this criterion. For the remainder of this paper we will
  work with this reduced sample of 112 pseudobulge-hosting S0
  galaxies.

The $D_n(4000)$ index distribution of reduced sample of
pseudobulge-hosting S0 galaxies is shown in left panel of Figure
\ref{fig:fig1}. One can clearly see a bimodality in $D_n(4000)$
distribution which translates primarily to the age bimodality in
pseudobulges of S0 galaxies. In order to systematically explore the
possible cause of this age bimodality, we have divided the sample of
pseudobulges in S0 galaxies with an old (having $D_n(4000)\geq 1.5$)
and a young (having $D_n(4000) < 1.5$) stellar population. A
  value of $D_n(4000)=1.5$ corresponds to a stellar age of $\sim$ 2
  Gyr \citep{Kauffmann2003}. The choice of the value $D_n(4000)=1.5$
to divide bulges into old and young types was done by examining the
left panel of Figure \ref{fig:fig1}. At this point, a clear dip in the
$D_n(4000)$ distribution is seen.  This dividing value has also been
used to select old and passive galaxies in recent literature
\citep{Zahid2017} \par

Pseudobulges are thought to be formed by transport of disc gas to the
central region of the galaxy \citep{Kormendy2016}. Since the amount of
gas in galaxies and star formation rate are correlated, one naively
expects that the age of the stellar population in the pseudobulges must be related to star formation rate of the galaxy as a whole. To understand this possible connection, we have
plotted our old and young pseudobulge host S0 galaxies on the
extinction corrected color-mass diagram. We have obtained total modelled
magnitude of S0 galaxies in $u$ and $r$ bands, and have corrected them
for extinction due to Galactic absorption by taking the extinction corrections in magnitudes (obtained following \citealt{Schlegel1998}) from the photoObj table given from SDSS DR13. We then have obtained K
corrected $u-r$ colour to $z = 0.0$ by making use of the the
K-corrections calculator
code\footnote{http://kcor.sai.msu.ru/getthecode/} which is based on
work by \cite{Chilingarian2010}. \par

 The extinction corrected color-stellar mass diagram for pseudobulge
 hosting S0 galaxies is shown in the middle panel of Figure
 \ref{fig:fig1}. The two solid lines, taken from
 \cite{Schawinski2014}, mark the boundary of the green valley region.
 The region above the green valley in this diagram is the passive red
 sequence and the region below the green valley is the star forming
 blue cloud. The S0 galaxies hosting old and young pseudobulges are
 shown by red and blue colours respectively. The median error in mass and colour estimate is about 0.15 dex and 0.015 mag respectively. We find that majority of
 old pseudobulges are hosted by passive S0 galaxies while young bulges
 are hosted by galaxies which are still forming stars in their
 disc. We also notice that some of the young pseudobulge-hosting S0
 galaxies are in the passive sequence, but one must be cautious here,
 as presence of dust can make galaxies redder and shift them towards
 the passive sequence of the $u-r$ color-mass diagram. The picture
 which emerges is then, of a connected history of star formation
 activity in the galaxy disc and in the the pseudobulges of S0
 galaxies. We surmise that the origin of the old population of
 pseudobulges in some S0 galaxies is due to shutting down of star
 formation in their disc. \par
 
 We have tried to explore the possible reason of shutting down of star
 formation leading to the age bimodality found in pseudobulges
 hosting S0 galaxies. Star formation in galaxies depends on
 morphology and environment, then it becomes worthwhile to check their
 possible correlation with age distribution of pseudobulges in S0
 galaxies. In the right panel of Figure \ref{fig:fig1}, we have plotted a 2D
 histogram of average environmental density vs $D_n(4000)$
 distribution for pseudobulges hosting S0 galaxies in our sample. We
 have obtained the average environmental density from \cite{Nair2010},
 which defines it as the logarithm of inverse of distance to 5'th
 nearest neighbour as defined in \cite{Baldry2006}. 
 
We see a weak trend of the environment with $D_n(4000)$ index
  where the old pseudobulge-hosting S0 appear to be in slightly higher
  environmental density as compared to the young ones. We have
  performed a two-sample Kolmogorov-Smirnov test to compare the
  environmental density distribution of young and old pseudobulges. We
  find that these samples of old and young pseudobulges could not have
  been drawn from the same parent population, with at least 99.8\%
  confidence. The mean environmental density of old and young
  pseudobulge-hosting S0s is -0.1 Mpc$^{-2}$ and -0.5 Mpc$^{-2}$
  respectively, although there is sufficient overlap in density
  parameter space which weakens the trend. This indicates that
  environment plays at most a weak role in driving the age bimodality
  of pseudobulges in S0 galaxies.

 \par To explore the importance of morphology of S0 galaxies in our
 sample, we have divided them into two morphological subclasses. In the 
 literature, one generally puts the morphological classes S0-, S0, S0+
 and S0/a galaxies under an umbrella term S0 galaxies
 \citep{Laurikainen2011}. Out of these S0/a are more closer to spirals
 and can have some faint spiral arms visible, while the others in S0
 class are mostly featureless and are closer to ellipticals
 galaxies. We have clubbed together the S0-, S0, S0+ morphological
 classes and call them as early-type S0 galaxies. In the left panel of
 Figure \ref{fig:both}, we have plotted the $D_n(4000)$ distribution
 for early-type S0 and S0/a galaxies. We notice that majority of old
 pseudobulges are found in early-type S0s while the younger ones are
 more common in S0/a galaxies, but when combined they form the peaks of
 bimodal distribution seen in the left panel of Figure \ref{fig:fig1}. We have also
 plotted the young and old pseudobulge-hosting early type S0 and S0/s
 galaxies on $u-r$ color-mass plane in the right panel of Figure
 \ref{fig:both}. We find the same trend between global star formation
 history and pseudobulge age as found previously in middle panel of Figure
 \ref{fig:fig1}, where old pseudobulges are associated with
 passive galaxies while the young bulges are predominantly found in
 star forming galaxies.  This plot then clearly shows that the age
 bimodality in pseudobulges of S0 galaxies is strongly driven by the
 morphology. We speculate that the morphology, which is shaped by the
 dynamical history of the galaxies, quenches the star formation in the
 disc, which then in turn stops the inward flow of gaseous material
 from the galaxy disc to the bulges, thus contributing to the ageing
 of pseudobulges as seen in these S0 galaxies.

\section{Summary}
\label{sec:sum}

We have presented a comparative study of the stellar
populations of pseudobulges hosted by S0 and spiral galaxies. We have
presented evidence of pseudobulge age bimodality in S0 galaxies which
is not seen in pseudobulges of spirals. Dividing the bulges into those
containing old and young populations, we see that old pseudobulges are
hosted by passive S0 galaxies, while the star forming S0 galaxies tend
to host young pseudobulges. We have tried to investigate the origin of
this age bimodality in pseudobulges of S0 galaxies by studying the
possible effect of the environment and the morphology. We do not see
any strong environmental effect which might drive this
bimodality. Dividing pseudobulge-hosting S0s into finer bins of
morphology, we find that early-type S0s preferentially host an older
pseudobulge while in the late-type S0s, i.e. the S0/a morphological
class, most of the pseudobulges are young. We surmise that the origin
of the old population of pseudobulges in some S0 galaxies is due to
quenching of star formation in their disc. We believe that the
dynamical history of these galaxies may have shaped their morphology
and may have quenched their disc, stopping the inward transport of
disc gas and thus making the bulges older by preventing the formation
of new stars. \par
 
 In future, we plan to investigate the stellar population of
 pseudobulges of S0 galaxies in detail using data from recent IFU
 surveys such as SDSS MANGA. By studying the star formation history of
 individual components of galaxies such as bulge, bar and the disc,
 one can get more insight on the connection between the pseudobulge
 stellar population and its relation to the star formation history of
 the disc.

\section*{Acknowledgements}

 We thank the anonymous referee for insightful comments that have improved both the content and presentation of this paper. We acknowledge support from a South African National Research Foundation grant (PID-93727) and from a bilateral grant under the Indo-South
Africa Science and Technology Cooperation (PID-102296) funded by the
Departments of Science and Technology (DST) of the Indian and South
African Governments.









\bsp	
\label{lastpage}
\end{document}